\documentclass[]{emulateapj}

\def\ni{\noindent}

\shorttitle{Cepheid P-L Relation with Testimator \& SIC}
\shortauthors{Kanbur et al.}

\begin{document}

\title{Investigations of the Non-Linear LMC Cepheid Period-Luminosity Relation with Testimator and Schwarz Information Criterion Methods}

\author{S. M. Kanbur}
\affil{Department of Physics, State University of New York at Oswego, Oswego, NY 13126}

\author{C. Ngeow}
\affil{Department of Astronomy, University of Illinois, Urbana, IL 61801}

\author{A. Nanthakumar and R. Stevens}
\affil{Department of Mathematics, State University of New York at Oswego, Oswego, NY 13126}

\begin{abstract}
In this paper, we investigate the linearity versus non-linearity of the Large Magellanic Cloud (LMC) Cepheid period-luminosity (P-L) relation using
two statistical approaches not previously applied to this problem: the testimator method and the Schwarz Information Criterion (SIC).
The testimator method is extended to multiple stages for the first time,
shown to be unbiased and the variance of the estimated slope can be proved to be smaller
than the standard slope estimated from linear regression theory.
The Schwarz Information Criterion (also known as the Bayesian Information Criterion) is more conservative than the Akaike Information
Criterion and tends to choose lower order models.
By using simulated data sets, we verify that these statistical techniques can be 
used to detect intrinsically linear and/or non-linear P-L relations. These methods are then applied to independent 
LMC Cepheid data sets from the OGLE project and the MACHO project, respectively. Our results imply that there is a change of slope 
in longer period ranges for all of the data sets. This strongly supports previous results, 
obtained from independent statistical tests, that the observed LMC P-L relation is non-linear with a break period at/around 10 days. 

\end{abstract}


\section{Introduction}

The cornerstone of the extra-galactic distance scale is the Cepheid Period-Luminosity (P-L) relation defined by the 
Large Magellanic Cloud (LMC) Cepheids. The assumed linear relation of the LMC Cepheid P-L relation, which is linear 
in $\log(P)$, with $P$ the pulsation period in days, has been under debate due to recent results that this 
relation could be non-linear \citep{tam02,kan04,san04,nge05}. These authors contended that the existing Cepheid data in the LMC 
strongly suggested the LMC P-L relation is consistent with two lines of significantly differing slopes with a break 
at/around a period of 10 days. This is referred as the non-linearity of the Cepheid P-L relation in this paper. 
Arguments for choosing the fiducial period at 10 days can be found in \citet{kan04}, \citet{san04}, \citet{nge05} and \citet{nge06a}. 
Furthermore, \citet{kan04,kan06}, \citet{san04}, \citet{nge05} and \citet{nge06} examined various factors that may cause the 
non-linearity of the LMC P-L relation, including the observing strategies, photometric errors, extinction errors,
removal of outliers, sample selection, number of long period Cepheids in the samples and
contamination of overtone Cepheids. They found that none of these remedies or any combination of them
could be responsible for the observed non-linear LMC P-L relation. As 
argued in \citet{nge06}, rigorous statistical tests are needed to test the linearity versus the non-linearity of the LMC 
P-L relation.

In our previous studies, the $F$-test \citep[e.g.][]{wei80} has been applied to the OGLE (Optical Gravitational Lensing 
Experiment) and MACHO (MAssive Compact Halo Objects project) Cepheid data, in \citet{kan04} and \citet{nge05} respectively, 
to test for the non-linearity of the LMC P-L relation. In such a formulation, the full and reduced models are models with 
four and two parameters respectively. This test looks at the change in the mean residual sum of squares between the full 
and reduced model divided by the mean residual sum of squares in the full model (see equation [5] of \citealt{kan04}). This test 
statistic can be formulated as the difference in slopes between short and long period slopes divided by the standard error 
of that difference. Hence if the number and nature of the long/short period data are such that the long/short period slope 
is estimated with a large error, then the $F$-value will be low and return a non-significant result. Thus the $F$-test 
is sensitive to the number of data points on either side of the period cut at 10 days. The OGLE and MACHO data sets 
we used in \citet{kan04} and \citet{nge05}, respectively, do have adequate number of long and short period Cepheids for 
the application of the $F$-test. The $F$-test has returned a significant result when testing the non-linearity of the P-L 
relation in both of the data sets.

Nevertheless, the results that suggesting a non-linear LMC P-L relation are still controversial. As we emphasize that statistical tests are needed, however claims of linear LMC P-L relation in the literature lack of rigorous statistical tests. In this paper, we apply two additional statistical tests, the testimator method and the Schwarz Information Criterion method, to examine the non-linearity of the LMC Cepheid P-L relation. These tests will be complementary to the $F$-test carried out in previous studies since they will serve to check and verify the results obtained from the $F$-test. In this way previous conclusions about the non-linear LMC P-L relation are considerably strengthened. Furthermore, both testimator and Schwarz Information Criterion methods are also able to estimate the break period without any a priori assumption: recall that in previous work, the break period at 10 days is usually adopted. These two methods not only can be applied for Cepheid studies, as we did in this paper, but also to other astronomical and astrophysical hypothesis testing problems. We also emphasize that our use of the testimator has, for the first time, been generalized to more than two stages and hence is also a statistical result in its own right. In the next section, we outline these techniques in detail and their application to our problems. In Section 3 we apply these methods to LMC Cepheid data and present our results. The conclusions and discussion are given in the last section.

\section{The Statistical Methods}

\subsection{The Testimator}

The concept of a testimator (or test estimator) was first proposed by \citet{ban44} in the context of estimating a parameter 
where a prior guess will be used in place of the estimator of an unknown parameter. The testimator can be applied if the prior 
guess for the unknown parameter can be ascertained by a test of hypothesis, otherwise the traditional estimator will be used. 
Due to its superior efficiency compared to traditional estimators, the testimator method has been adapted and refined to suit 
other situations by \citet{pau50}, \citet{hun55}, \citet{ban64}, \citet{arn72}, \citet{boc73}, \citet{han78}, 
\citet{gho88}, \citet{yan89}, \citet{pan90}, \citet{pan95}, \citet{pan97} and \citet{pan01} to name a few. \citet{wai84} and 
\citet{wai01}, in work on two-stage shrinkage estimation, proposed a weighted testimator by placing a weight $1-k$ on the prior 
guess and weight $k$ on the traditional estimator, where $k$ is the probability that the guess will be true. They showed that 
the testimators have far superior efficiency and therefore are more powerful in estimating unknown parameters. This weighted
two stage testimation concept can be extended to cover multiple stages. In this paper we apply this ``weighted'' testimator to 
investigate the non-linearity of the LMC Cepheid P-L relation as mentioned in the Introduction. 

The description of the two-stage testimator method is summarized as follows. 
For a linear regression of the form of $y={\beta} x + a$, the usual least square estimation of the slope to $N$ data points is 
given as

\begin{eqnarray}
\hat{\beta} & = & {{\sum_{i=1}^{N}(x_i-\bar{x})(y_i-\bar{y})}\over{\sum_{i=1}^{N}(x_i-\bar{x})^2}}, 
\end{eqnarray}

\ni where $\bar{x}=N^{-1}\sum x_i$ and $\bar{y}=N^{-1}\sum y_i$ are the mean values of $x$ and $y$, respectively. In the standard
hypothesis testing procedure, the null and alternate hypotheses are constructed as $H_0: {\beta} = \beta_0$ and 
$H_a: {\beta} \neq \beta_0$, respectively, where $\beta_0$ is the assumed value of (true) slope given the prior knowledge on the
slope. For example, $\beta_0$ can be predicted from theoretical calculations. In case that the (true) variance of the slope is known, 
the $z$-statistical test (with normal-distribution) can be applied, otherwise the $t$-statistical test (with $T$-distribution) will be 
used for the hypothesis testing. In general the variance is not known, therefore we adopt the $t$-statistical test in this paper.
If the null hypothesis is accepted from the hypothesis testing, the testimator (of the slope), $\hat{\beta_\omega}$, is calculated as 
\citep{wai84}:

\begin{eqnarray}
\hat{\beta}_{\omega} = k{\hat {\beta}} + (1-k){\beta_0}.
\end{eqnarray}

\ni The constant $k$ in the above equation is defined as 

\begin{eqnarray}
k & = & \frac{|t_{\mathrm observed}|}{t_{\mathrm critical}}, \\ 
t_{\mathrm observed} & = & \frac{\hat{\beta}-\beta_0}{\sqrt{MSE/S_{XX}}}, \nonumber \\
t_{\mathrm critical} & = & t_{\alpha/2,\nu}, \nonumber
\end{eqnarray}

\ni where $MSE=(N-2)^{-1}\sum^{N}_{i=1}(y_i-\hat{a}-\hat{\beta}x_i)^2$, $S_{XX}=\sum_{i=1}^{N}(x_i-\bar{x})^2$ and $t_{\alpha/2,\nu}$
is the $t$-value for $100(1-\alpha/2)$\% confidence interval obtained from the associated $T$-distribution table with $\nu=N-2$ degree of 
freedom. Note that the null hypothesis is rejected if $k>1$. The properties of the testimator are such that:

\begin{enumerate}
\item The testimator is an unbiased estimator under $H_0$. 
\item The testimator has a smaller variance than the usual least square estimator, that is $\mathrm{Var}(\hat{{\beta}_{\omega}}) < \mathrm{Var}(\hat{{\beta}})$. 
\end{enumerate}

\ni The proofs for these two properties are given in the Appendix.

\subsubsection{Application to the Cepheid P-L Relation}

\begin{figure}
\plotone{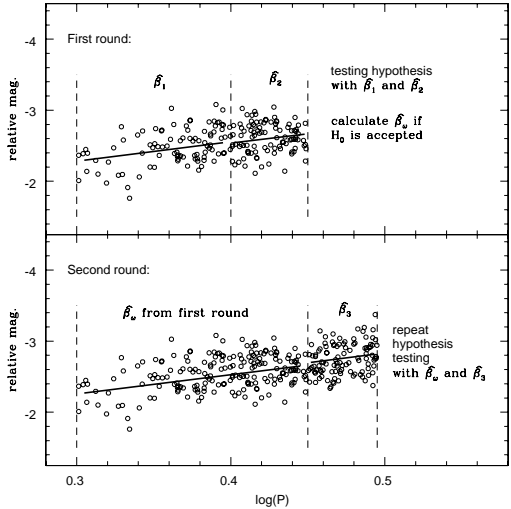}
\caption{Illustration of the testimator procedures. The Cepheid data points are divided to several subsets, sorted according to the $\log(P)$. For the first round, two slopes from the first and second subsets are compared under the hypothesis testing. The testimator, $\hat{\beta_\omega}$, is calculated if the null hypothesis is accepted. In the second round, the testimator from the previous round will be used to compare the estimated slope from the third subset. This is repeated several times until all the subsets have been used up or the null hypothesis is rejected. \label{testimator}}
\end{figure}

The motivation of this paper is to apply the testimator method to detect any non-linearity in the LMC P-L relation; this has been
detected using the $F$-test \citep{kan04,nge05}. To study any possible variation in slope as the period increases through
10 days, we first
sorted the data according to period, from shortest to longest period in $\log(P)$. The sorted sample is then divided into $m$ number of
non-overlapping and hence independent subsets according to the Cepheid period. The purpose is to make the bi-variate 
observations independent for each of the subsets. Each of the subsets will then contain $n$ numbers of Cepheids (if the number of data points in 
the last subset is small, then the last subset will be combined with the previous subset). This enables us to apply the testimator method in multiple
stages, together with a conservative Bonferroni testing procedure\footnote{The Bonferroni testing procedure states that
for testing $n_g$ number of hypotheses, 
the confidence coefficient $(1-\alpha/2)$ is replaced by $(1-\alpha/2n_g)$ in each of the hypothesis testings. This is to ensure that the overall 
confidence coefficient will not be less than the original desired value of $(1-\alpha/2)$.}, for detecting such a slope variation in the sample. 
In essence the line of attack is to 
compute the slope of the first subset and then compare with the slope of the next subset. If the two slopes are ``similar'', we look at the slope of the 
third subset with the smoothed slope obtained from a combination of all the previous subsets.
Hence, at the $i^{th}$ round, the slope of a given subset of the data is computed and compared 
with the smoothed slope from the testimator of all the previous data points. If the two slopes are statistically equivalent, then the current
subset of data will be incorporated into the computation of the smoothed slope and compared with the slope of the next subset of data. This smoothness 
is an important feature since it helps to alleviate, to some extent, the influence of outliers.  However, if the slopes are ``different'', i.e. a 
rejection of the null hypothesis, then there is an indication of slope change in the P-L relation. Therefore there will be a total of $n_g=m-1$ number 
of hypothesis testings in the multi-stage testimator procedures. 
In short, the algorithm of applying  the testimator method in our case can be summarized as follows:

\begin{itemize}
\item[a.] In the first round, the slope of first subset, ${\beta}_1$, is calculated and denoted by $\hat{\beta}_1=\beta_0$. The slope of the second subset is then compared to ${\beta}_0$ under the null hypothesis of $H_0: {\beta}_2 = {\beta}_0$ and alternate hypothesis of $H_a: {\beta}_2\neq {\beta}_0$. If $H_0$ is accepted, then the testimator in this round, $\hat{{\beta}_{\omega}}^1$, is calculated using equation (2).

\item[b.] In the second round, the slope of the third subset, ${\beta}_3$, is calculated and denoted by $\hat{\beta}_3$. The testimator from the first round, represented as $\hat{{\beta}_{\omega}}^1=\beta_0$, is used in the hypothesis testing for this round. 
The null and alternate hypotheses in this round become $H_0: {\beta}_3 ={\beta}_0$ and $H_a:{\beta}_3 \neq {\beta}_0$.
If $H_0$ is accepted, a new testimator, $\hat{\beta_\omega}^2$, is calculated using equation (2).

\item[c.] In the $i^{th}$ round, the slope of the $(i+1)^{th}$ subset, ${\beta}_{i+1}$, estimated by $\hat{\beta}_{i+1}$, is calculated. The testimator from previous ($i-1$) round is denoted as $\hat{{\beta}_{\omega}}^{i-1}=\beta_0$. The null and alternative hypothesis in this round become $H_0:{\beta}_{i+1} = {\beta}_0$ and $H_a:{\beta}_{i+1} \neq {\beta}_0$. If $H_0$ is accepted, then $\hat{\beta_\omega}^i=k\hat{\beta}_{i+1}+(1-k)\beta_0$ with $k$ refined from equation (3). 

\item[d.] This is repeated until $i=n_g$ round or the null hypothesis is rejected in the $i^{th}$ round, which indicates a change in slope for the $(i+1)^{th}$ subset. 

\item[e.] Since in principle there will be a total of $n_g$ hypothesis testings, the Bonferroni testing procedure requires that 
$t_{\mathrm critical}=t_{\alpha/2n_g,\nu}$ in each round.
 
\end{itemize}

\ni Throughout the paper, we will adopt $\alpha=0.05$ to ensure the overall confidence level is more than 95\% in our test. 
The first two rounds of our testimator procedures to the study the possible non-linear LMC P-L relation is illustrated in Figure \ref{testimator}.

\begin{deluxetable*}{cccccccccc}
\tabletypesize{\scriptsize}
\tablecaption{Testimator results for the fake data sets.\label{tab1}}
\tablewidth{0pt}
\tablehead{
\colhead{Subset} & \colhead{Period range} & \colhead{$n$} & \colhead{$\hat{\beta}$} & \colhead{$\beta_0$} 
& \colhead{$|t_{\mathrm observed}|$} & \colhead{$t_{\mathrm critical}$} 
& \colhead{$k$} & \colhead{Decision} & \colhead{$\hat{\beta_{\omega}}$} \\
\colhead{(1)} & \colhead{(2)} & \colhead{(3)} & \colhead{(4)} 
& \colhead{(5)} & \colhead{(6)} & \colhead{(7)} & \colhead{(8)} & \colhead{(9)}
& \colhead{(10)}
}
\startdata
\multicolumn{10}{c}{``Fake'' data set from a linear P-L relation} \\
1 & 0.2315-0.4421 & 200 & $-2.182\pm0.403$ & ---    & ---   & ---   &  ---  & ---          & ---    \\
2 & 0.4422-0.5005 & 200 & $-3.658\pm0.949$ & $-2.182$ & 1.556 & 2.718 & 0.572 & accept $H_0$ & $-3.027$ \\
3 & 0.5006-0.5508 & 200 & $-1.955\pm1.128$ & $-3.027$ & 0.951 & 2.718 & 0.350 & accept $H_0$ & $-2.652$ \\
4 & 0.5512-0.6079 & 200 & $-3.006\pm1.025$ & $-2.652$ & 0.345 & 2.718 & 0.127 & accept $H_0$ & $-2.697$ \\
5 & 0.6080-0.7349 & 200 & $-2.733\pm0.442$ & $-2.697$ & 0.081 & 2.718 & 0.030 & accept $H_0$ & $-2.698$ \\
6 & 0.7349-0.9610 & 200 & $-2.841\pm0.234$ & $-2.698$ & 0.611 & 2.718 & 0.225 & accept $H_0$ & $-2.730$ \\
7 & 0.9610-1.3553 & 200 & $-2.493\pm0.155$ & $-2.730$ & 1.531 & 2.718 & 0.563 & accept $H_0$ & $-2.597$ \\
8 & 1.3652-2.6170 & 100 & $-2.684\pm0.095$ & $-2.597$ & 0.921 & 2.748 & 0.335 & accept $H_0$ & $-2.626$ \\
\multicolumn{10}{c}{``Fake'' data set from a non-linear P-L relation} \\
1 & 0.2315-0.4421 & 200 & $-2.442\pm0.403$ & ---    & ---   & ---   &  ---  & ---          & ---    \\
2 & 0.4422-0.5005 & 200 & $-3.918\pm0.949$ & $-2.442$ & 1.556 & 2.718 & 0.572 & accept $H_0$ & $-3.287$ \\
3 & 0.5006-0.5508 & 200 & $-2.215\pm1.128$ & $-3.287$ & 0.951 & 2.718 & 0.350 & accept $H_0$ & $-2.912$ \\
4 & 0.5512-0.6079 & 200 & $-3.266\pm1.025$ & $-2.912$ & 0.345 & 2.718 & 0.127 & accept $H_0$ & $-2.957$ \\
5 & 0.6080-0.7349 & 200 & $-2.993\pm0.442$ & $-2.957$ & 0.081 & 2.718 & 0.030 & accept $H_0$ & $-2.958$ \\
6 & 0.7349-0.9610 & 200 & $-3.101\pm0.234$ & $-2.958$ & 0.611 & 2.718 & 0.225 & accept $H_0$ & $-2.990$ \\
7 & 0.9610-1.3553 & 200 & $-2.170\pm0.155$ & $-2.990$ & 5.281 & 2.718 & 1.943 & reject $H_0$ & ---    
\enddata
\tablecomments{See text for the description for each columns. Period ranges are given in $\log(P)$.}
\end{deluxetable*}

In order to demonstrate the reliability of this procedure, we apply the testimator method to two simulated data sets: one built from a non-linear P-L 
relation with a break at 10 days and another one developed from a linear P-L relation. For demonstration purpose, we select one set of the simulated data (out of many simulations) in each cases as representation for testing the testimator method.
The plots of these two fake data sets, each of them containing 1500 data points, can be found in figure 1 of \citet{nge06}. 
Full details of our procedure for developing these two ``fake'' 
data sets are described in \citet{nge06}. The results of applying the testimator procedures as described to these two fake data sets are given in Table \ref{tab1}. 
In this table, column 1 denotes the subsets; column 2 gives the range of the period in each subsets; column 3 lists the number of data points, $n$,
in each subsets; column 4 \& 5 are the fitted slopes in each subsets and the assigned values of $\beta_0$ that used in the hypothesis testing; column 6 \& 7 are 
the observed and critical $t$-values for each of the hypothesis testing; column 8 \& 9 are the corresponding $k$-value and the outcome of the 
hypothesis testing; finally column 10 is the values of testimator if the null hypothesis is accepted. Since we know which fake data set is intrinsic 
linear and non-linear when constructing the P-L relation, we can verify the results found in Table \ref{tab1}. For the fake data with linear P-L
relation, our testimator results show that the slopes for each subsets are consistent with the smoothed slopes given from the previous subsets, and
the hypothesis testings correctly indicate that there is no changes in slope across all the period ranges. In contrast, the hypothesis testings for the 
fake data with non-linear P-L relation show that subset 7 has a different slopes than the previous subsets, which indicates a change of slope in this 
subset. Furthermore, the testimator procedures also correctly pick up the ``break period'' in subset 7, which brackets
the input break period at 10 days, from the outcome of hypothesis testing. Therefore the testimator method can pick up the P-L relation which is 
intrinsically non-linear.

\subsection{The Schwarz Information Criterion}

The problem of deciding whether the LMC Cepheid data are more consistent with two lines of significantly different slopes rather than a 
single line is exactly analogous to deciding the dimensionality of the model that will fit the given LMC Cepheid data.
The method of maximizing the likelihood tends to choose the highest possible dimension. \citet{aka74} suggested maximizing the likelihood
subject to a penalty depending on the dimensionality of the model under consideration (Akaike Information Criterion, AIC): $AIC=-2\ln L + 2k_p$,
where $L$ is the likelihood function of the model of dimension $k_p$ \citep[see][as an example in the application of astronomy]{tak00}. However, \citet{sch78} showed that maximum likelihood
estimators can be obtained from large sample limits of Bayes estimates for certain classes of a priori distributions. These distributions only
put positive probability on the subspaces of the parameter space corresponding to the competing models. \citet{sch78} derived the following
criterion (Schwarz Information Criterion, SIC; or sometimes also referred as Bayesian Information Criterion, BIC, in the literature): choose the model for which 

\begin{eqnarray}
SIC = -2\ln L + k_p\ln N
\end{eqnarray}

\ni is a minimum, where $N$ is the total number of data points and $k_p=p+1$ \citep[with $p$ being the number of fitted parameters, see][]{sch78}. Some use of the BIC for models selection in astronomical and astrophysical literature can be found, for examples, in \citet{are01}, \citet{han00,han02}, \citet{koe96,koen99,koe06}, \citet{koe99}, \citet{koe00}, \citet{koe93,koe03}, \citet{lid04,lid07}, \citet{muk98}, \citet{por06} and \citet{ste99}.

\subsubsection{Application to the Cepheid P-L relation}

To test the non-linearity of the Cepheid P-L relation with the SIC method, we consider the models with a linear P-L relation (the null hypothesis) and a non-linear P-L relation with a break period (in days) at $P_0$ (the alternate hypothesis) in this paper. For the former case, we have:

\begin{eqnarray}
H_0: m = \hat{m} & = & \hat{\beta} \log(P) + \hat{a},  \nonumber \\
 & &   \mathrm{with} \ \hat{\sigma}^2=\frac{1}{N-2}\sum_{i=1}^{i=N} (m_i-\hat{m}_i),  \nonumber \\
L & = & \frac{1}{(\sqrt{2\pi \hat{\sigma}^2})^N} \exp{[-\frac{1}{2\hat{\sigma}^2} \sum^{N}_{i=1} (m_i - \hat{m_i})^2]}, \nonumber \\
SIC(H_0) & = & -2\ln L + 3\ln N. \nonumber
\end{eqnarray}

\ni Similarly, for the alternate model, we have:

\begin{eqnarray}
H_A: m = \hat{m}  & = &\left\{  \begin{array}{l}
                   \hat{\beta}_S \log(P) + \hat{a}_S, \ \mathrm{for} \log(P) < \log(P_0),  \\
                     \ \ \ \ \mathrm{with}\ \hat{\sigma}_S^2=\frac{1}{N_S-2}\sum_{i=1}^{i=N_S} (m_i-\hat{m}_i),  \\
                   \hat{\beta}_L \log(P) + \hat{a}_L, \ \mathrm{for} \log(P) \geq \log(P_0),  \\
                     \ \ \ \ \mathrm{with}\ \hat{\sigma}_L^2=\frac{1}{N_L-2}\sum_{i=1}^{i=N_L} (m_i-\hat{m}_i),
                  \end{array}\right. \nonumber 
\end{eqnarray}
\begin{eqnarray}
L & = & \frac{1}{(\sqrt{2\pi})^N}\frac{1}{(\hat{\sigma}_S)^{N_S}}\frac{1}{(\hat{\sigma}_L)^{N_L}} \nonumber \\ 
& & \exp{[-\frac{1}{2\hat{\sigma}_S^2} \sum^{N_S}_{i=1} (m_i - \hat{m_i})^2 -\frac{1}{2\hat{\sigma}_L^2} \sum^{N_L}_{i=1} (m_i - \hat{m_i})^2]}, \nonumber 
\end{eqnarray}
\begin{eqnarray}
SIC(H_A)  =  -2\ln L + 5\ln N. \nonumber
\end{eqnarray}

\ni In these expressions, $N_S+N_L=N$ and $m$ is the observed magnitudes after correcting for extinction. The slope ($\beta$) and zero-point ($a$) parameters in the above models are obtained from the maximum likelihood estimation (MLE, which is equivalent to standard least square estimation in our case). Note that the sample variance ($\sigma^2$) from MLE is a biased estimate. We corrected the bias with $N_{(L,S)}-2$ degrees of freedom. For the alternate models, $SIC(H_A)$ is calculated with a range of $\log(P_0)$ that increment in steps of, for example, $0.001$. Therefore, a model with linear P-L relation and a range of models with non-linear P-L relations at different break period are tested with the SIC method. The model with smallest $SIC$ value is the preferred model. In case of $SIC(H_A)<SIC(H_0)$,  the minimum value of $SIC(H_A)$ not only suggests that the P-L relation is non-linear, but also gives an estimate of the break period. 

\begin{figure*}
\plottwo{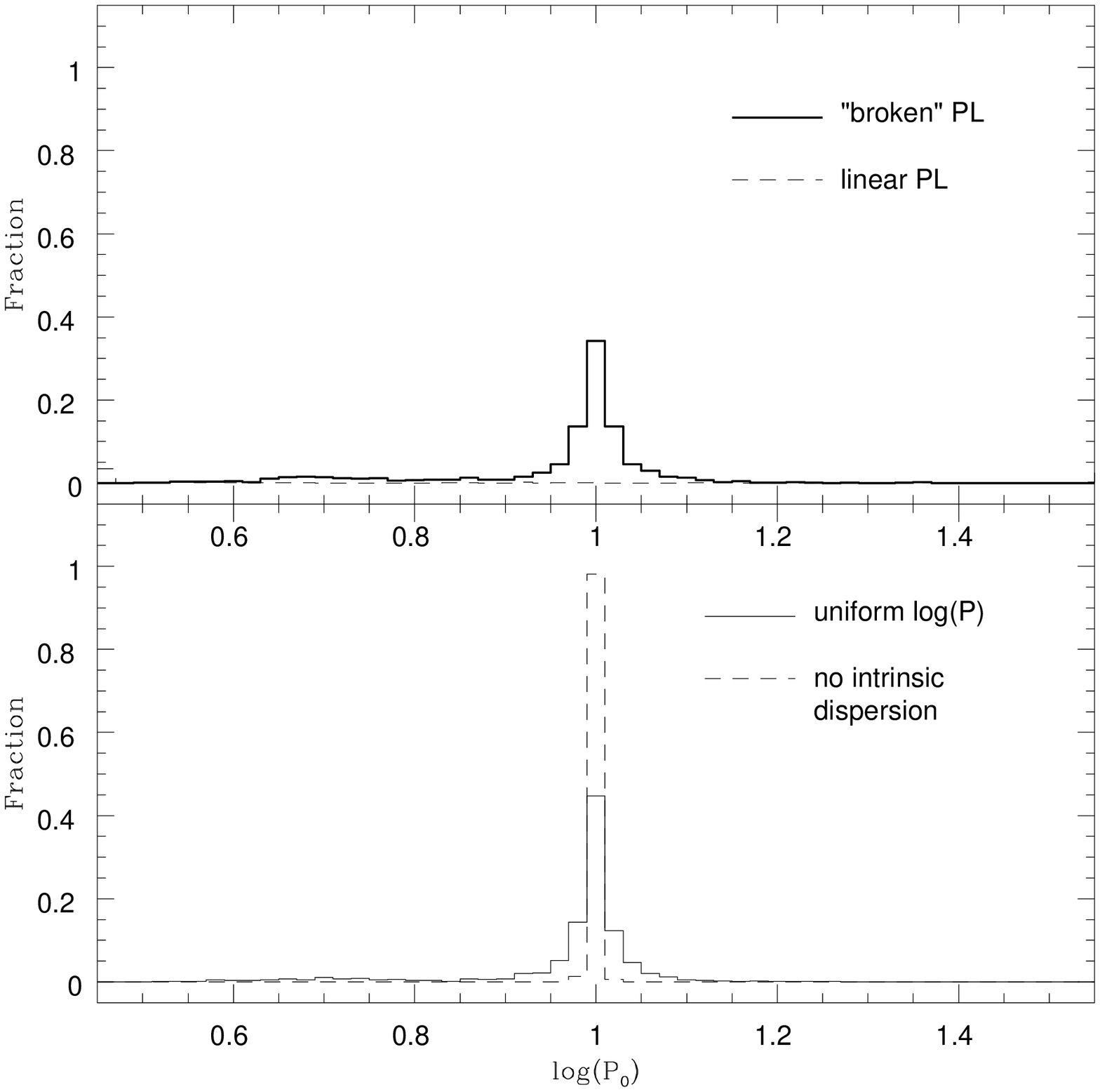}{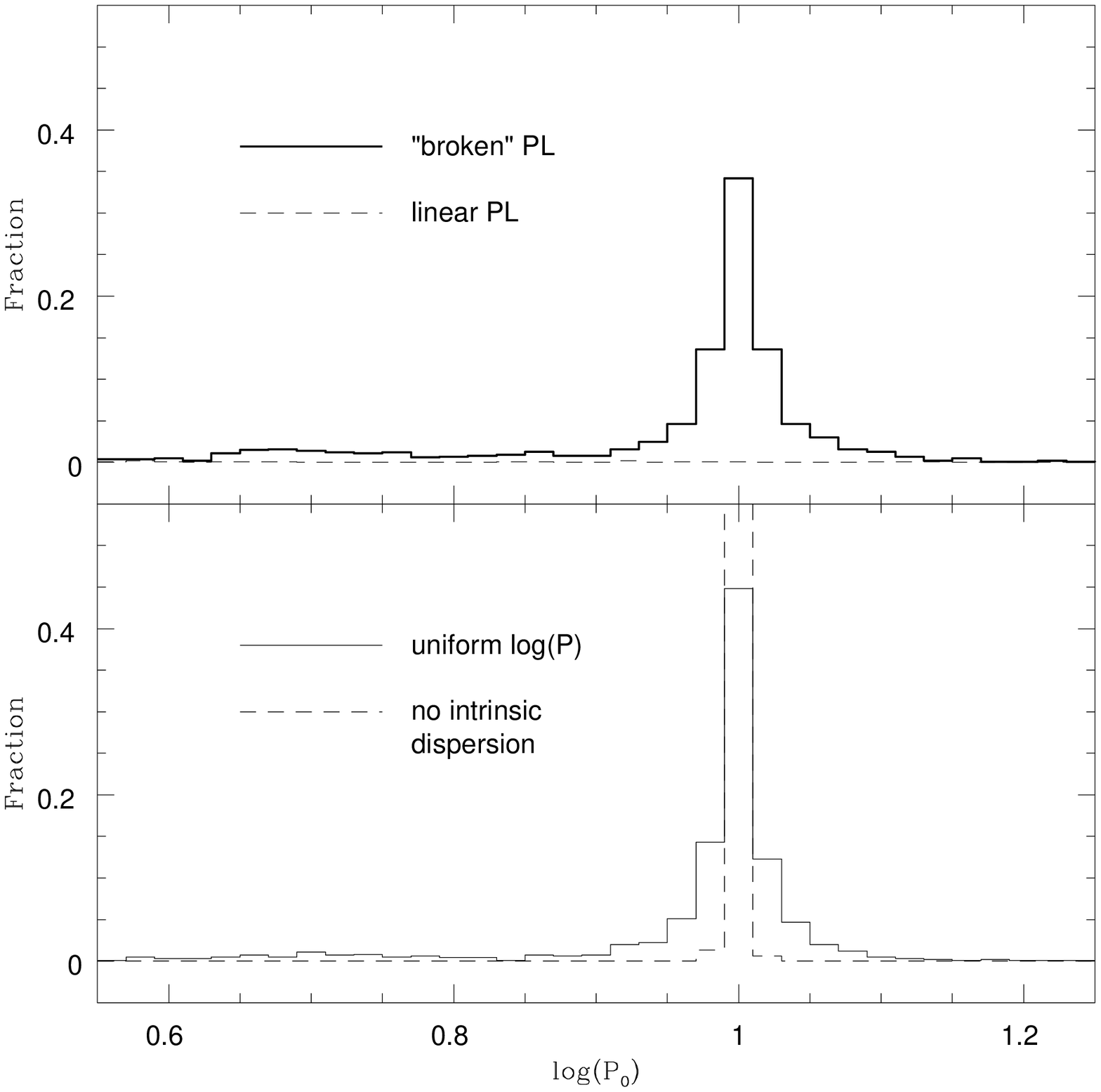}
\caption{Distributions of the estimated break periods, $\log(P_0)$, using the SIC method with various simulations. The top panels show the histograms from two simulations at which the input P-L relations to the simulation is linear (dashed histogram) and non-linear with a break period at $\log(P_0)=1.0$ (thick histogram), respectively. The bottom panels show the histograms from two additional simulations using the same non-linear P-L relation with a break period again at $\log(P_0)=1.0$ as input P-L relation: one simulation has uniform distribution of the periods (in $\log[P]$) in the simulated data, and another simulation did not include the intrinsic dispersion of the P-L relation. The right panels show the blown-up region of the left panels. \label{simu_dist}}
\end{figure*}

To test the SIC method, the same simulated data sets as in the case of the testimator method in Section 2.1.1 were used. For the ``fake'' data set a with linear P-L relation, the values of $SIC(H_0)$ and $SIC(H_A)$ is $-164.65$ and $-161.21$, respectively. While for the ``fake'' data set with non-linear P-L relation, we found $SIC(H_0)=-100.56$ and the minimum value of $SIC(H_A)=-154.62$ occurs at $\log(P_0)=0.983$. We then test the SIC method for our application further with various simulations. We first ran two sets of simulations: one set of simulations use the linear P-L relation as input P-L relation, and another set of simulations include the non-linear P-L relation with a break at $\log(P_0)=1.0$. These simulations mimic the period distribution and the observed dispersions along the P-L relation from the real data. The details for constructing these simulations can be found in \citet{nge06}. In either sets of the simulations, a large number of simulations is run (typically 1000) and the break period (in $\log[P_0]$) is searched with the SIC method. If the break period cannot be found then this implies the linear P-L relation is the preferred model, and vice versa. The top panels of Figure \ref{simu_dist} display the distributions of the break periods from these two sets of data. For the case of a linear P-L relation, the SIC method did not find any break period $\sim90$\% of the time. While for the case of no-linear P-L relation, the SIC method detects a range of break period with a peak at $\log(P_0)\sim1.0$. Therefore, the SIC method can be used to correctly identify the P-L relation that is either intrinsically linear or non-linear at a given break period. 

The relatively large dispersion around $\log(P_0)\sim1.0$ and the long tail toward shorter period that is exhibited in the top panels of Figure \ref{simu_dist} can be due to a combination of two effects: (1) the existence of the intrinsic dispersion along the P-L relation; and (2) the non-uniform distribution of the periods in the data (see \citealt{nge05} and \citealt{nge06} for more discussion about the period distribution for Cepheid variables). To portray the impact of these effects on the application of the SIC method for detecting the break period, we ran two additional experiments. The first retains the original period distribution but the intrinsic dispersion of the P-L relation is excluded (however the random photometric errors still persist in the simulation), while the second simulation use a uniform period distribution (in $\log[P]$) and the intrinsic dispersion of the P-L relation is not excluded. The resulted distributions of the break period from SIC method are presented in the bottom panels of Figure \ref{simu_dist}. It can be seen that the long tail of the distribution present in the top panels is reduced when a uniform period distribution is assumed. Furthermore, if the intrinsic dispersion does not exist in the Cepheid P-L relation, then the SIC method is very efficient to detect the intrinsic break period (at $\log[P_0]=1.0$ in our case). In reality, the intrinsic dispersion along the P-L relation cannot be eliminated or reduced (at least in the optical bands) and the period distribution of the Cepheid variables will not be uniform (for the reasons discussed in \citealt{nge06}). We emphasize that the theoretical pulsation modelings are needed to identify the location of the break period or to confirmation the break period at $\log(P_0)\sim1.0$ \citep{nge05}.

\section{Data and Results}

In this section, we apply both the testimator and SIC methods to the real LMC Cepheids data in order to investigate whether the $V$-band Cepheid P-L 
relation at {\it mean light} is non-linear or not. We concentrate on the $V$-band mean light data because the data 
available in the literature are mostly in the $V$-band mean light and also because of the evidence for non-linearity as a function of phase is clear 
\citep{nge06a}. The data sets we used in this study include the OGLE data adopted from \citet{kan06} and the 
MACHO data adopted from \citet{nge05}. They are referred as the ``OGLE'' sample (with 641 Cepheids) and the ``MACHO'' data (with 1216 Cepheids), 
respectively. Note that both data sets have been corrected for extinction using the method described in the corresponding papers. It
is also important to point out that these two are independent data sets. To investigate the influence of longer period stars in our testing
as well as increasing the number of Cepheids in the OGLE sample, we append the data from \citet{seb02} to the OGLE sample after proper removal of 
duplication of the Cepheids in both samples and the correction of extinction. This third data set is called ``OGLE+SEBO'' sample (with 723 Cepheids) and 
it extends to $\log(P)\sim2.0$.

\begin{deluxetable*}{cccccccccc}
\tabletypesize{\scriptsize}
\tablecaption{Testimator results for the real data sets.\label{tab2}}
\tablewidth{0pt}
\tablehead{
\colhead{Subset} & \colhead{Period range} & \colhead{$n$} & \colhead{$\hat{\beta}$} & \colhead{$\beta_0$} 
& \colhead{$|t_{\mathrm observed}|$} & \colhead{$t_{\mathrm critical}$} 
& \colhead{$k$} & \colhead{Decision} & \colhead{$\hat{\beta_{\omega}}$} \\
\colhead{(1)} & \colhead{(2)} & \colhead{(3)} & \colhead{(4)} 
& \colhead{(5)} & \colhead{(6)} & \colhead{(7)} & \colhead{(8)} & \colhead{(9)}
& \colhead{(10)}
}
\startdata
\multicolumn{10}{c}{OGLE sample, Test 1} \\
1 & 0.4022-0.4771 & 100 & $-1.427\pm0.967$ & ---    & ---   & ---   &  ---  & ---          & ---    \\
2 & 0.4787-0.5293 & 100 & $-2.273\pm1.399$ & $-1.427$ & 0.605 & 2.627 & 0.230 & accept $H_0$ & $-1.622$ \\
3 & 0.5294-0.5889 & 100 & $-0.746\pm1.095$ & $-1.622$ & 0.800 & 2.627 & 0.304 & accept $H_0$ & $-1.355$ \\
4 & 0.5891-0.6703 & 100 & $-1.887\pm0.675$ & $-1.355$ & 0.788 & 2.627 & 0.300 & accept $H_0$ & $-1.515$ \\
5 & 0.6704-0.7891 & 100 & $-3.055\pm0.703$ & $-1.515$ & 2.193 & 2.627 & 0.835 & accept $H_0$ & $-2.801$ \\
6 & 0.7900-1.6768 & 141 & $-2.462\pm0.082$ & $-2.801$ & 4.106 & 2.612 & 1.572 & reject $H_0$ & --- \\
\multicolumn{10}{c}{OGLE sample, Test 2} \\
1 & 0.4022-0.5043 & 150 & $-2.547\pm0.647$ & ---    & ---   & ---   &  ---  & ---          & ---    \\
2 & 0.5043-0.5889 & 150 & $-1.783\pm0.641$ & $-2.547$ & 1.193 & 2.421 & 0.493 & accept $H_0$ & $-2.171$ \\
3 & 0.5891-0.7083 & 150 & $-2.347\pm0.401$ & $-2.171$ & 0.438 & 2.421 & 0.181 & accept $H_0$ & $-2.203$ \\
4 & 0.7103-1.6768 & 191 & $-2.590\pm0.075$ & $-2.203$ & 5.139 & 2.415 & 2.128 & reject $H_0$ & --- \\
\multicolumn{10}{c}{OGLE+SEBO sample, Test 1} \\
1 & 0.4022-0.4746 & 100 & $-0.989\pm0.882$ & ---    & ---   & ---   &  ---  & ---          & ---    \\
2 & 0.4752-0.5242 & 100 & $-2.476\pm1.202$ & $-0.989$ & 1.237 & 2.693 & 0.459 & accept $H_0$ & $-1.672$ \\
3 & 0.5245-0.5729 & 100 & $-4.743\pm1.339$ & $-1.672$ & 2.292 & 2.693 & 0.851 & accept $H_0$ & $-4.286$ \\
4 & 0.5734-0.6469 & 100 & $-2.743\pm0.907$ & $-4.286$ & 1.701 & 2.693 & 0.632 & accept $H_0$ & $-3.311$ \\
5 & 0.6491-0.7320 & 100 & $-2.921\pm0.933$ & $-3.311$ & 0.418 & 2.693 & 0.155 & accept $H_0$ & $-3.250$ \\
6 & 0.7330-0.9071 & 100 & $-3.315\pm0.400$ & $-3.250$ & 0.162 & 2.693 & 0.060 & accept $H_0$ & $-3.254$ \\
7 & 0.9112-2.1268 & 123 & $-2.497\pm0.089$ & $-3.254$ & 8.535 & 2.682 & 3.181 & reject $H_0$ & --- \\
\multicolumn{10}{c}{OGLE+SEBO, Test 2} \\
1 & 0.4022-0.4977 & 150 & $-2.545\pm0.546$ & ---    & ---   & ---   &  ---  & ---          & ---    \\
2 & 0.4891-0.5729 & 150 & $-2.826\pm0.706$ & $-2.545$ & 0.398 & 2.529 & 0.157 & accept $H_0$ & $-2.589$ \\
3 & 0.5734-0.6831 & 150 & $-2.557\pm0.432$ & $-2.589$ & 0.073 & 2.529 & 0.029 & accept $H_0$ & $-2.588$ \\
4 & 0.6831-0.9071 & 150 & $-3.153\pm0.253$ & $-2.588$ & 2.234 & 2.529 & 0.883 & accept $H_0$ & $-3.087$ \\
5 & 0.9112-2.1268 & 123 & $-2.497\pm0.089$ & $-3.087$ & 6.651 & 2.536 & 2.623 & reject $H_0$ & --- \\
\multicolumn{10}{c}{MACHO sample} \\
1 & 0.4008-0.4715 & 200 & $-2.391\pm0.958$ & ---    & ---   & ---   &  ---  & ---          & ---    \\
2 & 0.4719-0.5226 & 200 & $-1.843\pm1.189$ & $-2.391$ & 0.462 & 2.601 & 0.178 & accept $H_0$ & $-2.294$ \\
3 & 0.5231-0.5787 & 200 & $-2.623\pm1.127$ & $-2.294$ & 0.292 & 2.601 & 0.112 & accept $H_0$ & $-2.331$ \\
4 & 0.5795-0.6851 & 200 & $-1.851\pm0.809$ & $-2.331$ & 0.594 & 2.601 & 0.228 & accept $H_0$ & $-2.222$ \\
5 & 0.6588-0.7891 & 200 & $-2.948\pm0.524$ & $-2.222$ & 1.385 & 2.601 & 0.533 & accept $H_0$ & $-2.608$ \\
6 & 0.7910-1.4501 & 216 & $-2.123\pm0.122$ & $-2.608$ & 3.991 & 2.599 & 1.536 & reject $H_0$ & --- 
\enddata
\end{deluxetable*}

The results from using the testimator method to these three LMC Cepheid data sets are summarized in Table \ref{tab2}, with identical layout as Table \ref{tab1}.
In the case for the OGLE and OGLE+SEBO data sets, we have tried different sample subset sizes
by dividing the
samples to $n=100$ and $n=150$, which are referred as Test 1 and Test 2 in the table, respectively. In all cases, the testimator method implies that there 
is a change of slope in the last subset of the samples. Similar results found from Test 1 and Test 2 suggest that our results are not affected by the 
size of each subset. This indicates the LMC P-L relation becomes non-linear as the period increases through 10 days to longer periods. 
The last subset also brackets the fiducial break period at/around 10 days: this is consistent with previous work of \citet{nge05}.

\begin{figure}
\plotone{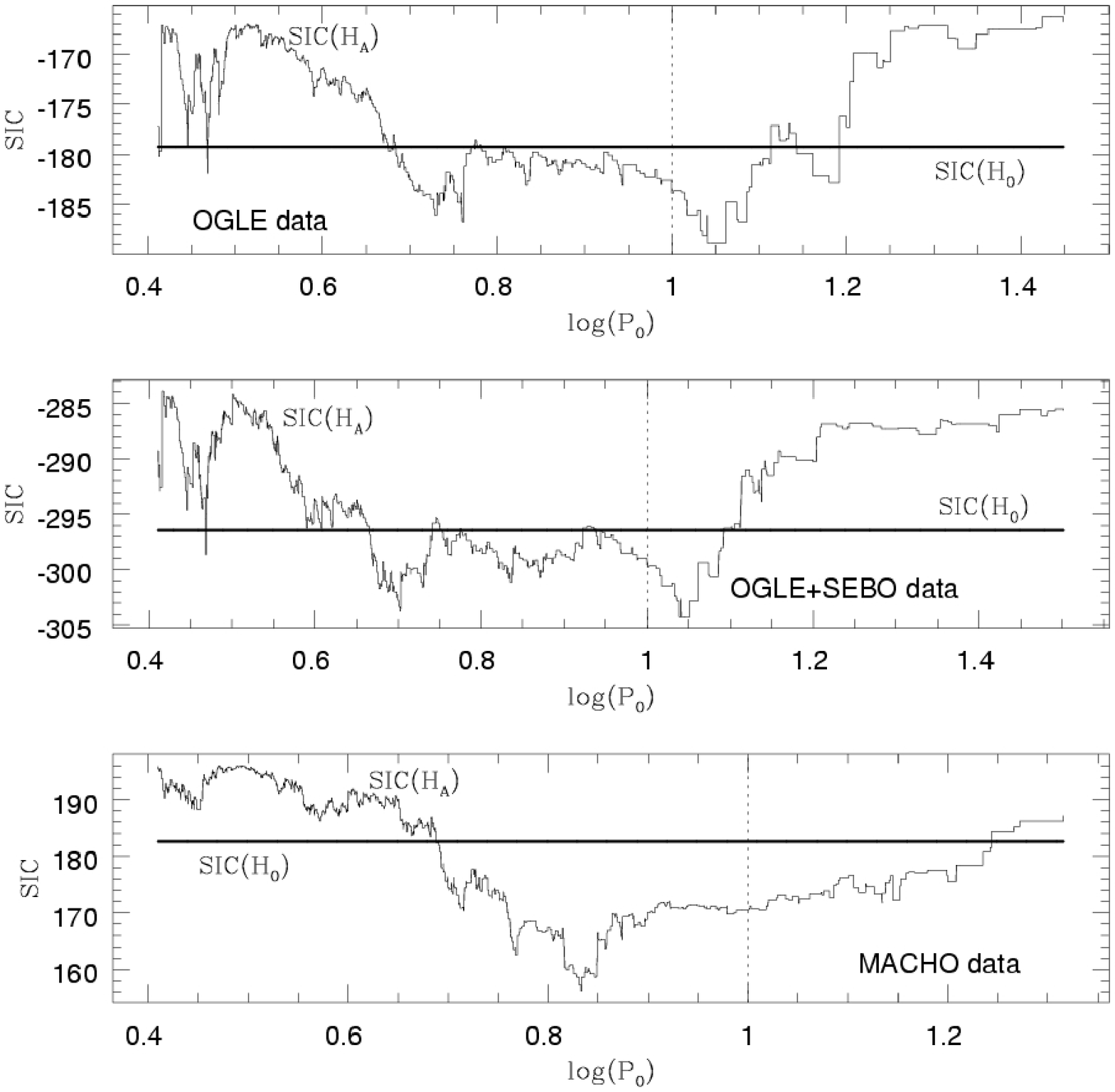}
\caption{Results of the $SIC(H_0)$ and $SIC(H_A)$ as a function of the choosing break period, $\log(P_0)$, for the three LMC Cepheid data sets. The thick horizontal lines are the results for $SIC(H_0)$, which are independent of the chosen break period. The ``curves'' are the results for $SIC(H_A)$. The horizontal dotted lines represent the chosen break period in the literature \citep[e.g.,][]{tam02,kan04,san04,nge05}. \label{sic_result}}
\end{figure}

\begin{figure}
\plotone{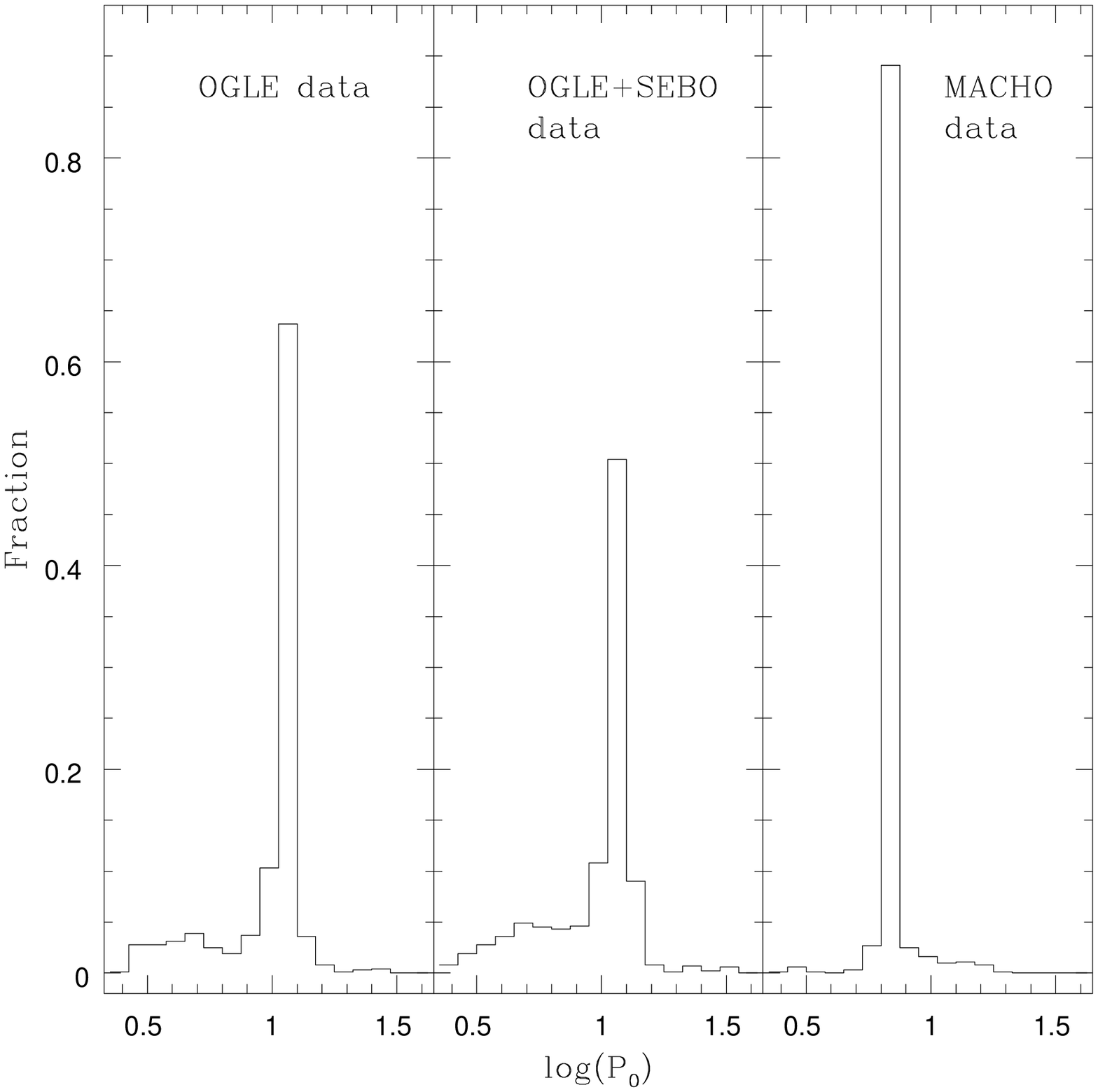}
\caption{Resulted histograms from the bootstrap re-sampling at the break period given in Table \ref{tab3} for the three LMC Cepheid data sets. See text for details. \label{sic_hist}}
\end{figure}

\begin{figure}
\plotone{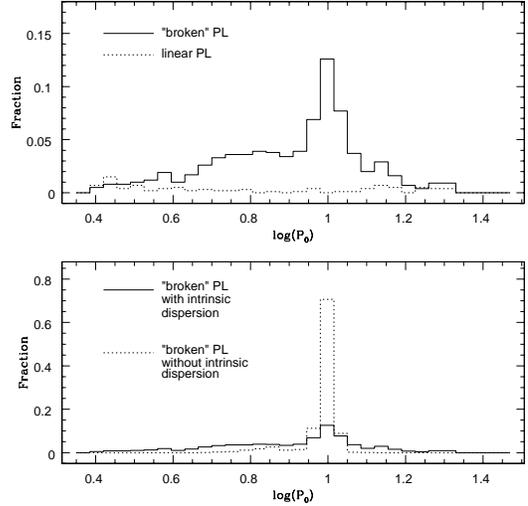}
\caption{Comparisons of the histograms from three sets of simulations for the MACHO data: (1) a simulation that takes a non-linear P-L relation with a break at $\log(P_0)=1.0$ as input P-L relation and the intrinsic dispersion is included; (2) a simulation with a linear P-L relation as input P-L relation and the intrinsic dispersion is included; and (3) a simulation that takes a non-linear P-L relation with a break at $\log(P_0)=1.0$ as input P-L relation but without the intrinsic dispersion. Unlike other simulations done in this paper, the periods that go into the simulations are from the {\it actual} MACHO data.  \label{macho}}
\end{figure}

The results from using the SIC method are presented in Figure \ref{sic_result} and Table \ref{tab3} for the same data sets. In Figure \ref{sic_result}, the values of SIC for both $SIC(H_0)$ and $SIC(H_A)$ are plotted as a function of the chosen break period, $\log(P_0)$. Since $SIC(H_0)$ is independent of $\log(P_0)$, this represents a straight horizontal line in the figure, and the values of $SIC(H_0)$ for these three data sets are given in Table \ref{tab3}. For the case of $SIC(H_A)$ as a function of $\log(P_0)$, the figure bears witness to the fact that there is a range of $\log(P_0)$ at which the values of $SIC(H_A)$ are smaller than $SIC(H_0)$ in all three data sets. This implies that the non-linear P-L relation is preferred within these period ranges. This result also reinforces the findings of Figure \ref{simu_dist} that it is difficult to determine the exact break period of the P-L relation with SIC method (see Section 2.2.1 as well), if it is present. The minimum values for $SIC(H_A)$ found from the figure, and the corresponding $\log(P_0)$ are summarized in Table \ref{tab3} as well. The confidence intervals for the break period can be estimated using bootstrap re-sampling methods. For the model with given $\log(P_0)$ in  Table \ref{tab3}, the errors of the regression, $\epsilon_i=m_i-\hat{m}_i$, are randomly drawn (with replacement) to construct a ``new'' data set, and a new break period is estimated. This is repeated many times to build up the distribution of the break periods. The resulting histograms for these three sets of data are presented in Figure \ref{sic_hist}. From these distributions, the $5^{\mathrm{th}}$-, $25^{\mathrm{th}}$-, $75^{\mathrm{th}}$- and $95^{\mathrm{th}}$-percentile are estimated for each of the data sets. The results are given in the last four columns of Table \ref{tab3}. At first glance the break period found from the MACHO data seems to be inconsistent with the OGLE and OGLE+SEBO results. This is due to the difficulty of accurately estimating the break period with the existence of the instability strip. To demonstrate this, we use the {\it exact} periods in MACHO data as input periods to our simulations, and generate three different sets of simulations: (1) a simulation with intrinsic non-linear P-L relation; (2) a simulation with linear P-L relation; and (3) a simulation with intrinsic non-linear P-L relation but without the intrinsic dispersion. The resulting histograms for these three sets of simulation are displayed in Figure \ref{macho}. From this figure it is clear that our result of the break period for MACHO data does not imply an inconsistency to the OGLE and OGLE+SEBO results. The break period found in the data, $\log(P_0)=0.833$, is within the range of the break periods found from the simulations. This figure also portray the difficulty of estimating the break period from real data when the intrinsic dispersion along the P-L relation is present. Therefore, the break periods given in Table \ref{tab3} are consistent with the results from testimator (Table \ref{tab2}), the result from non-linear estimation procedure applied in \citet[][$\log(P_0)=0.934$ with upper and lower 95\% confidence level of 1.089 and 0.778, respectively]{nge05} and the adopted $\log(P_0)=1.0$ in the literature. Note that in previous studies \citep[e.g.,][]{tam02,kan04,kan06,san04,nge05,nge06c} the break period is conveniently chosen to be at 10 days, which is represented as dotted vertical line in Figure \ref{sic_result}. The SIC results also supported the non-linear P-L relation to be the preferred model at $\log(P_0)=1.0$. 

\begin{deluxetable*}{lccccccc}
\tabletypesize{\scriptsize}
\tablecaption{SIC results for the real data sets.\label{tab3}}
\tablewidth{0pt}
\tablehead{
\colhead{Dataset} & \colhead{$SIC(H_0)$} & \colhead{$SIC(H_A)$} & \colhead{$\log(P_0)$} 
& \colhead{$5^{\mathrm{th}}$-percentile} & \colhead{$25^{\mathrm{th}}$-percentile}
& \colhead{$75^{\mathrm{th}}$-percentile} & \colhead{$95^{\mathrm{th}}$-percentile}
}
\startdata
OGLE sample                 & $-179.29$ & $-188.86$ & 1.041 & 0.550 & 1.002 & 1.041 & 1.101 \\
OGLE+SEBO sample            & $-296.43$ & $-304.28$ & 1.041 & 0.560 & 0.922 & 1.052 & 1.131 \\
MACHO sample                & $182.61$  & $156.20$  & 0.833 & 0.806 & 0.826 & 0.838 & 0.936 
\enddata
\end{deluxetable*}

\section{Conclusion and Discussion}

Using two additional statistical approaches, the method of testimators and SIC, to the OGLE, the OGLE+SEBO and the MACHO Cepheids data, we have found strong statistical 
evidence for a change of slope in the Cepheid P-L relation in the LMC at longer period range. These results also strongly support the previous results 
obtained from the $F$-test. Therefore, the {\it observed} LMC P-L relation is non-linear based on these rigorous statistical tests. This implies that either
the LMC P-L relation is indeed non-linear or there are some hidden factors in the analysis \citep[see][for more discussion on this]{nge06}. Furthermore, the break periods, or the range of permissible break periods found from this study are consistent with the conveniently chosen break period at 10 days in previous studies. However, our study, both with real and fake data, implies that it is difficult to accurately estimate the break period with both the testimator and  SIC methods. This is due to the existence of the intrinsic dispersion along the P-L relation. The confirmation of the break period at/around 10 days has to be done from stellar pulsation modeling studies. 

The implications 
of a non-linear LMC P-L relation on the extra-galactic distance scale and the Hubble constant have been discussed in \citet{nge05w} and \citet{nge06c} and will 
not be repeated here.
A number of authors, including \citet{spe06}, \citet{teg06}, \citet{mac06}, \citet{oll06} and the reference therein, have commented on how an independent estimate of 
the Hubble constant accurate to $1\%$ can significantly reduce the error bars on ${\Omega}$, the total density of the universe. Applying the correct form of the 
Cepheid P-L relation will help in reducing the systematic error of the Hubble constant in the future studies \citep{nge06c,nge06}. 
Over and above this, if the 
Cepheid P-L relation does indeed have a change of slope at 10 days, it is important to understand this from a stellar
pulsation and evolution point of view 
and investigate fully the ramifications of this new feature \citep{kan06,mar05}. 

\citet{nge06} has investigated various factors that may cause the observed non-linear LMC  P-L relation, including the influence of outliers and lack of
longer period Cepheids in the sample. However the results from that study suggest that none of these factors are
responsible for the observed non-linear LMC P-L relation.
We emphasize that the samples we used in the our studies have been cleaned up for obvious outliers. Further, the testimator approach estimates the slope 
with a variance which is smaller than the standard formula (property 2 stated in Section 2.1.1) is able to minimize the effect of (additional) outliers by smoothing.
Regarding the lacks of longer period Cepheids in the sample, we have use the OGLE+SEBO as an expansion sample to the OGLE sample with the increase 
of period coverage. Both of the samples have shown the same testimator ans SIC results. Therefore, we believe this should not be the cause for the observed 
non-linear LMC P-L relation. 

\acknowledgements

The authors would like to thank the referee for useful suggestions. This research was supported in part by NASA through the American Astronomical Society's Small Research Grant Program. 

\appendix
\section{Proof for the Properties of the Testimator}

We prove the two properties of the testimator as described in Section 2.1 here. To prove that the testimator is an unbiased estimator under 
$H_0$, we note that the testimator from equation (1) is:

$$\hat{\beta}_{\omega} = k({\hat{\beta}} - {\beta}_0) + {\beta}_0,$$

\ni where $k$ is defined in equation (3). Therefore the above expression can be re-written as:

\begin{eqnarray}
\hat{\beta}_{\omega} = {{|\hat{\beta} - {\beta}_0|(\hat{\beta} - {\beta}_0)}\over{t_{{\alpha}/2,\nu} \sqrt{MSE/S_{XX}}}} + \beta_0.
\end{eqnarray}

\ni This implies that,

$$E({\hat{{\beta}_{\omega}}}) = {\sqrt{S_{XX}}\over{t_{{\alpha}/2,\nu}}}E({{1}\over{\sqrt{MSE}}}) E[|{\hat{{\beta}}} - {\beta}_0|({\hat{{\beta}}}-{\beta}_0)] + \beta_0.$$

\ni Since $E(|z|z)=0$ for variable $z=\hat{\beta} - \beta_0$ with standard normal distribution, and from the above expression, we obtain

$$E({\hat{{\beta}_{\omega}}}) = {\beta}_0$$

\ni as desired. The second assertion states that $\mathrm{Var}(\hat{{\beta}_{\omega}}) < \mathrm{Var}(\hat{{\beta}})$. To proof this, we first re-arrange equation (A1) such that:

$$(\hat{\beta}_{\omega} - \beta_0)^2 = \frac{(\hat{\beta} - \beta_0)^4}{t^2_{\alpha/2,\nu}\ MSE} S_{XX}.$$

\ni Assume $\hat{\beta}$ is normally distributed with $N(\beta_0,\sigma_\beta)$ and define $Z=\frac{\hat{\beta} - {\beta}_0}{\sigma_{\hat{\beta}}}$, 
then $Z$ has a standard
normal distribution with $N(0,1)$. Note that $\sigma^2_{\hat{\beta}}\equiv \mathrm{Var}(\hat{\beta})=\sigma^2/S_{XX}$, where $\sigma^2$ is the variance of the 
linear regression $y={\beta} x + a$, the above expression is reduced to:

$$ (\hat{\beta}_{\omega} - \beta_0)^2 = Z^4\frac{\sigma^2}{MSE} \frac{1}{t^2_{\alpha/2,\nu}}\frac{\sigma^2}{S_{XX}}.$$

\ni Hence, we have 

\begin{eqnarray}
Var(\hat{\beta}_{\omega}) & = & E(Z^4) E(\frac{\sigma^2}{MSE}) \frac{1}{t^2_{\alpha/2,\nu}} \frac{\sigma^2}{S_{XX}},
\end{eqnarray}

\ni as the last two terms are constants and $\mathrm{Var}(\hat{\beta}_{\omega})\equiv E[(\hat{\beta}_{\omega} - \beta_0)^2]$. For $E(Z^4)$, since the
forth moment of a standard normal distribution (the Kurtosis) is 3, then $E(Z^4)=3$. For $E(\frac{\sigma^2}{MSE})$, we observe that 
$\frac{\sigma^2}{MSE}=\frac{1}{MSE/\sigma^2}=(N-2)/\sum(\frac{y_i-\hat{a}-\hat{\beta}x_i}{\sigma})^2$. Therefore, $(N-2)MSE/\sigma^2$ is $\chi^2$ distributed 
with $\nu=N-2$ degree of freedom. It is well-known that if $X$ is $\chi^2$ distributed with $\nu$ degree of freedom, then $E(1/X)=1/(\nu-2)$, hence 
$E(\frac{\sigma^2}{MSE})=(N-2)/(N-4)$. Recall that $\sigma^2/S_{XX}=\mathrm{Var}(\hat{\beta})$, then equation (A2) is reduced to:  

$$ \mathrm{Var}(\hat{\beta}_{\omega}) = 3\frac{(N-2)}{N-4}\frac{1}{t^2_{\alpha/2,\nu}} \mathrm{Var}(\hat{\beta}).$$

\ni If $t_{\alpha/2,\nu}>\sqrt{3(N-2)/(N-4)}$, we then have

$$\mathrm{Var}(\hat{{\beta}_{\omega}}) < \mathrm{Var}(\hat{{\beta}})$$

\ni as the assertion states. Due to the Bonferroni testing procedure, condition $t_{\alpha/2n_g,\nu}>\sqrt{3(N-2)/(N-4)}$ is satisfied when $N>5$ and ${\alpha}<0.1$.

\end{document}